\documentclass[prd,nofootinbib,superscriptaddress,onecolumn,preprintnumbers]{revtex4}
\usepackage{graphicx}
\usepackage{amsmath,amssymb}
\renewcommand{\eqref}[1]{Eq.(\ref{#1})}

\usepackage{color}

\begin{document}
\title{Decay of $f(R)$ quintessence into dark matter: mitigating the Hubble tension?}

\author{Giovanni Montani}
\email{giovanni.montani@enea.it}
\affiliation{Nuclear Department, ENEA - C. R. Frascati, Via E. Fermi 45, 00044 Frascati, Italy}
\affiliation{Physics Department, ``Sapienza'' University of Rome,  P.le Aldo Moro 5, 00185 Roma, Italy}

\author{Luis A. Escamilla}
\affiliation{Department of Physics, Istanbul Technical University, 34469 Maslak, Istanbul, Turkey}
\affiliation{School of Mathematics and Statistics, University of Sheffield, Hounsfield Road, Sheffield S3 7RH, United Kingdom}

\author{Nakia Carlevaro}
\affiliation{Nuclear Department, ENEA - C. R. Frascati, Via E. Fermi 45, 00044 Frascati, Italy}

\author{Eleonora Di Valentino}
\affiliation{School of Mathematics and Statistics, University of Sheffield, Hounsfield Road, Sheffield S3 7RH, United Kingdom}

\begin{abstract}
We propose a revised cosmological scenario that extends the $\Lambda$ Cold Dark Matter ($\Lambda$CDM) framework by incorporating metric $f(R)$ gravity in the Jordan frame. In this model, the dark energy component arises from a non-minimally coupled scalar field, decomposed into a smooth background (set to unity to recover General Relativity) and a rapidly varying, massive fluctuation that decays into the dark matter sector. In the near-GR limit, this setup provides a phenomenological extension of $\Lambda$CDM characterized by two additional parameters: the present-day value of the scalar fluctuation and a normalized decay rate.  
Using a Markov Chain Monte Carlo analysis of low-redshift cosmological data, comprising Type Ia Supernovae, Baryon Acoustic Oscillation (BAO), and Cosmic Chronometer measurements, we find that the proposed model achieves a better overall fit than $\Lambda$CDM, while the Bayesian evidence remains statistically inconclusive given the inclusion of two extra parameters. The model predicts a moderate increase in the inferred value of $H_0$ and an improved consistency with DESI BAO data when adopting the SH0ES prior. Furthermore, describing dark matter particle creation as a transition phase in the late Universe offers an intriguing physical interpretation, potentially capturing features already present in current data and providing a promising avenue to explore extensions of the standard cosmological model within modified gravity frameworks.

\end{abstract}

\maketitle

\section{Introduction}
The discrepancy between the inferred values of the Hubble constant $H_0$ as measured by SH0ES collaboration~\cite{Riess:2021jrx,Breuval:2024lsv} and by Cosmic Microwave Background (CMB) experiments~\cite{Planck:2018vyg,Planck:2018nkj,ACT:2020gnv,ACT:2025fju,SPT-3G:2025bzu} has attracted increasing attention in recent years. This discrepancy, commonly referred to as the \emph{Hubble tension}~\cite{Verde:2019ivm,DiValentino:2020zio,DiValentino:2021izs,Perivolaropoulos:2021jda,Schoneberg:2021qvd,Shah:2021onj,Abdalla:2022yfr,DiValentino:2022fjm,Kamionkowski:2022pkx,Giare:2023xoc,Hu:2023jqc,Verde:2023lmm,DiValentino:2024yew,Perivolaropoulos:2024yxv,CosmoVerse:2025txj}, remains unexplained by any known astrophysical effect~\cite{Dainotti:2021vyp,Efstathiou:2020wxn,Mortsell:2021nzg,Mortsell:2021tcx,Riess:2021jrx,Sharon:2023ioz,Murakami:2023xuy,Riess:2023bfx,Bhardwaj:2023mau,Brout:2023wol,Dwomoh:2023bro,Uddin:2023iob,Riess:2024ohe,Freedman:2024eph,Riess:2024vfa}. This motivates the search for new dynamical features of the Universe that could account for the observed discrepancy~\cite{DiValentino:2021izs}.
Two main lines of research have emerged in the literature~\cite{CosmoVerse:2025txj}. On the one hand, late-Universe modifications of the Hubble expansion rate have been proposed~\cite{Dutta:2018vmq,vonMarttens:2019ixw,DiValentino:2020naf,DiValentino:2020vnx,Yang:2021flj,DiValentino:2021rjj,Heisenberg:2022lob,Giare:2023xoc,Adil:2023exv,Gomez-Valent:2023uof,Lapi:2023plb,Krolewski:2024jwj,Bousis:2024rnb,Jiang:2024xnu,Manoharan:2024thb,DiValentino:2017oaw,DiValentino:2017iww,DiValentino:2017rcr,Akarsu:2019hmw,Akarsu:2021fol,Akarsu:2022typ,Schiavone:2022wvq,Montani:2023ywn,Montani:2023xpd,Escamilla:2024xmz,Giare:2024ytc,Escamilla:2023shf,Montani:2025jkk,Dainotti:2021pqg,Dainotti:2022bzg,Fazzari:2025mww,Specogna:2025guo,Sabogal:2025mkp,Teixeira:2024qmw,DiValentino:2019exe,Alestas:2021luu,Anchordoqui:2023woo,Ruchika:2023ugh,Frion:2023xwq,Ruchika:2024ymt}. On the other hand, early-Universe modifications have been invoked to alter the size of the acoustic horizon, thereby affecting the CMB-inferred value of $H_0$~\cite{Kamionkowski:2022pkx,Poulin:2018cxd,Smith:2019ihp,Niedermann:2019olb,Krishnan:2020obg,Schoneberg:2021qvd,Ye:2021iwa,Poulin:2021bjr,Niedermann:2021vgd,deSouza:2023sqp,Poulin:2023lkg,Cruz:2023lmn,Niedermann:2023ssr,Efstathiou:2023fbn,Garny:2024ums,Giare:2024akf,Giare:2024syw,Poulin:2024ken,Pedrotti:2024kpn,Lynch:2024hzh,Toda:2024ncp,Schoneberg:2024ynd}. 
Beyond these two main categories, several studies have emphasized the potential need for a combined approach to fully and convincingly address the tension~\cite{Vagnozzi:2023nrq,Vagnozzi:2019ezj,Anchordoqui:2021gji,Jia:2022ycc}. An additional key contribution has recently come from Dark Energy Spectroscopic Instrument (DESI) Collaboration, whose analysis suggests that an evolving dark energy term provides a better fit to observations than a pure cosmological constant~\cite{DESI:2024mwx,DESI:2025zgx} (see also~\cite{Colgain:2024xqj,Cortes:2024lgw,Shlivko:2024llw,Luongo:2024fww,Yin:2024hba,Gialamas:2024lyw,Dinda:2024kjf,Najafi:2024qzm,Wang:2024dka,Ye:2024ywg,Tada:2024znt,Carloni:2024zpl,Chan-GyungPark:2024mlx,DESI:2024kob,Bhattacharya:2024hep,Ramadan:2024kmn,Notari:2024rti,Orchard:2024bve,Hernandez-Almada:2024ost,Pourojaghi:2024tmw,Giare:2024gpk,Reboucas:2024smm,Giare:2024ocw,Chan-GyungPark:2024brx,Menci:2024hop,Li:2024qus,Li:2024hrv,Notari:2024zmi,Gao:2024ily,Fikri:2024klc,Jiang:2024xnu,Zheng:2024qzi,Gomez-Valent:2024ejh,RoyChoudhury:2024wri,Li:2025cxn,Lewis:2024cqj,Wolf:2025jlc,Shajib:2025tpd,Giare:2025pzu,Chaussidon:2025npr,Kessler:2025kju,Pang:2025lvh,Roy:2024kni,RoyChoudhury:2025dhe,Paliathanasis:2025cuc,Scherer:2025esj,Giare:2024oil,Liu:2025mub,Teixeira:2025czm,Santos:2025wiv,Specogna:2025guo,Sabogal:2025jbo,Cheng:2025lod,Herold:2025hkb,Cheng:2025hug,Ozulker:2025ehg,Lee:2025pzo,Ormondroyd:2025iaf,Silva:2025twg,Ishak:2025cay,Fazzari:2025lzd}).

Among the various late-Universe approaches proposed to address the Hubble tension, one particularly appealing avenue involves \emph{modified gravity}, and in particular the metric $f(R)$ theories of gravity~\cite{Nojiri:2010wj,Schiavone:2022wvq,Schiavone:2024heb,Montani:2023xpd,Montani:2024xys}. In the so-called \emph{Jordan frame}~\cite{Sotiriou:2008rp}, these models introduce a non-minimally coupled scalar field in addition to the standard gravitational degrees of freedom. The functional form of the considered $f(R)$ theory is encoded in the potential term that governs the self-interaction of this scalar field.
Following the general idea of a dark energy--dark matter interaction discussed in~\cite{Montani:2024pou} (see also~\cite{Wang:2016lxa,Montani:2025rcy,Kumar:2016zpg,Murgia:2016ccp,Kumar:2017dnp,DiValentino:2017iww,Kumar:2021eev,Gao:2021xnk,Pan:2023mie,Benisty:2024lmj,Yang:2020uga,Forconi:2023hsj,Pourtsidou:2016ico,DiValentino:2020vnx,DiValentino:2020leo,Nunes:2021zzi,Yang:2018uae,vonMarttens:2019ixw,Lucca:2020zjb,Gao:2022ahg,Zhai:2023yny,Bernui:2023byc,Hoerning:2023hks,Giare:2024ytc,Escamilla:2023shf,vanderWesthuizen:2023hcl,Silva:2024ift,DiValentino:2019ffd,Li:2024qso,Pooya:2024wsq,Halder:2024uao,Castello:2023zjr,Yao:2023jau,Mishra:2023ueo,Nunes:2016dlj,Silva:2025hxw,Yang:2025uyv,vanderWesthuizen:2025rip,vanderWesthuizen:2025vcb,vanderWesthuizen:2025mnw}), we identify dark energy with the evolution of the non-minimally coupled scalar field. We then assume that scalar particles, interpreted as linear perturbations of the full scalar theory, can decay into dark matter particles. The interaction between the two species is governed by a phenomenological decay rate, which regulates the corresponding transfer of energy from the scalar sector to the dark matter component. 
We first derive the exact dynamical equations governing the dark energy--dark matter interaction and then implement a simplified phenomenological description of the proposed scenario. This simplified picture relies on two main hypotheses: (i) we consider a decay rate much greater than the Hubble constant, thereby treating the dark matter creation as a very late and rapid process, and (ii) we study the case in which the modified gravity theory arises from a linearly perturbed scalar field, i.e., we investigate the limit where the dominant background scalar field exhibits a frozen dynamics near the General Relativity (GR) regime. 
Once the dynamics of the modified Hubble parameter is obtained in this simplified framework, we perform a Markov Chain Monte Carlo (MCMC) analysis of the model using all available low-redshift datasets in the interval $0 \le z \le 2.5$. This analysis allows us to determine the most probable values of the four free parameters—two additional ones with respect to the standard $\Lambda$ Cold Dark Matter ($\Lambda$CDM) model. The resulting parameter constraints enable us to assess the capability of the proposed scenario to alleviate the Hubble tension.

The paper is organized as follows. In Sec.~\ref{sec00}, we introduce the Jordan frame representation of the metric $f(R)$ gravity. In Sec.~\ref{sec2}, we set up the cosmological dynamics of a flat, isotropic Universe in the considered $f(R)$ gravity, deriving the modified Friedmann, acceleration, and scalar field equations that govern the model. In Sec.~\ref{sec3}, we present our proposed framework, in which the scalar field is decomposed into a background component and a rapidly varying fluctuation that can decay into dark matter, thus generating an interaction between dark energy and dark matter. In Sec.~\ref{sec4}, we simplify the model into a reduced, dimensionless form introducing normalized parameters that capture the new dynamics. In Sec.\ref{sec5}, we discuss the near-GR limit of the model and the equivalence between Einstein and Jordan frames. In Sec.~\ref{sec5bis}, we describe the data analysis methodology, detailing the statistical tools, priors, and cosmological datasets used to constrain the model and compare it against $\Lambda$CDM. In Sec.~\ref{sec6}, we present and discuss the results, showing that our modified gravity scenario provides a modest improvement over $\Lambda$CDM in fitting the data and partially alleviates the Hubble tension. Concluding remarks follow.

\section{Metric $f(R)$-gravity}\label{sec00}

The most natural extension of GR that preserves all fundamental principles, affecting only the gravitational field dynamics, is the so-called ``metric $f(R)$-gravity''~\cite{Sotiriou:2008rp}, associated with the action:
\begin{equation}
	S_g = \frac{1}{2\chi}\int d^4x \sqrt{-g}f(R)\, , 
	\label{prdn1}
\end{equation}
where $\chi$ is the Einstein constant (we are in $c=1$ units), $g$ is the metric determinant, $R$ is the standard Ricci scalar, as calculated by the metric tensor $g_{\mu\nu}$ ($\mu ,\nu = 0,1,2,3$), while $f$ is a generic function. When the gravitational field does not live in vacuum, we should add to the action above the standard matter contribution, whose variation provides the Landau-Lifshitz energy-momentum tensor $T_{\mu\nu}^{(m)}$~\cite{landau2}. Hence, the field equations for 
a metric $f(R)$-gravity take the following fourth-order form in the metric tensor derivatives:
\begin{equation}
	(df/dR)R_{\mu\nu}-\frac{1}{2}fg_{\mu\nu} - \nabla_{\mu}\nabla_{\nu}(df/dR)+ 
	g_{\mu\nu}\nabla_{\rho}\nabla^{\rho}(df/dR)=\chi T^{(m)}_{\mu\nu}\, , 
	\label{prdn2}
\end{equation}
where $R_{\mu\nu}$ denotes the standard Ricci tensor and $\nabla_{\mu}$ is the Riemannian covariant derivative. Clearly, for $f\equiv R$, we recover the ordinary Einstein equations. Now, a well-known theorem of analytical mechanics and field theory states that, if given degrees of freedom can be eliminated from the field equations via the remaining ones, then they can be removed from the action by such a relation and the dynamics is preserved~\cite{Teitelboim}. Hence, introducing the auxiliary field $\theta$, the action in Eq.~(\ref{prdn1}) can be equivalently restated as:
\begin{equation}
	S_g = \frac{1}{2\chi}
	\int d^4x\sqrt{-g}[(df/dR)(R-\theta) + f]
	\, .
	\label{prdn3}
\end{equation}
In fact, if $(d^2f/dR^2)\neq 0$, then the variation of the action above with respect to $\theta$ yields $\theta = R$, which, once substituted back, makes Eq.~(\ref{prdn3}) coincide with the original action in Eq.~(\ref{prdn1}).

Introducing the scalar field $\phi \equiv df(\theta)/dR$, Eq.~(\ref{prdn3}) stands as follows:
\begin{equation}
	S_g = \frac{1}{2\chi}\int d^4x \sqrt{-g}[\phi R - V(\phi )]\, ,
	\label{prdn4}
\end{equation}
where
\begin{equation}
	V(\phi ) = \theta (df/dR) - f\, ,
	\label{prdn5}
\end{equation}
and the expression $\theta = \theta (\phi)$ comes from the (required) invertible definition of $\phi$ (here we can equivalently use $\theta$ or $R$, according to the literature~\cite{Sotiriou:2008rp}). It is important to stress how the request $(df/dR)>0$ ensures the non-repulsive character of gravity, while restricting the theory to the case $(d^2f/dR^2)\ge 0$ prevents the emergence of tachyon modes in the dynamics. 

Varying the action in Eq.~(\ref{prdn4}) with respect to the metric tensor yields the following field equations, which generalize the Einstein ones:
\begin{equation}
	\phi G_{\mu\nu}=\chi T^{(m)}_{\mu\nu} - \frac{1}{2}Vg_{\mu\nu} + \nabla_{\mu}\nabla_{\nu}\phi - g_{\mu\nu}\nabla_{\rho}\nabla^{\rho}\phi 
	\, , 
	\label{prdn6}
\end{equation}
where $G_{\mu\nu}$ denotes the Einstein tensor. We see how the equations above directly map into Eq.~(\ref{prdn2}) as soon as we substitute the relation $\phi=df/dR$ and Eq.~(\ref{prdn5}). This corresponds to the so-called ``Jordan frame'' representation and it is equivalent to the original theory (\ref{prdn1}), but it allows us to avoid the treatment of fourth-order field equations. However, the variation of Eq.~(\ref{prdn4}) with respect to $\phi$ naturally provides an additional field equation of the form
\begin{equation}
	R = \frac{dV}{d\phi}
	\, ,
	\label{prdn7}
\end{equation}
Combining together the trace of Eq.~(\ref{prdn6}) with Eq.~(\ref{prdn7}), we get a Klein-Gordon-like equation for the scalar field $\phi$, having the form
\begin{equation}
	3\nabla_{\rho}\nabla^{\rho}\phi + 2V - \phi \frac{dV}{d\phi} = \chi T_{\rho}^{\rho}
	\, .
	\label{prdn8}
\end{equation}

The formulation of the metric $f(R)$-gravity in the Jordan frame has the important advantage to deal with second-order field equations, like GR, but the emerging scalar-tensor theory, which is clearly a sub-case of the Brans–Dicke theory, is characterized by a non-minimal coupling between the gravitational and scalar fields. This problem could be addressed by passing to the so-called ``Einstein framework'' representation, based on the following conformal transformation:
\begin{equation}
	g_{\mu\nu} \equiv \tilde{g}_{\mu\nu}/\phi \equiv 
	\exp \left[-\sqrt{\frac{2\chi}{3}}\tilde{\phi}\right]\tilde{g}_{\mu\nu}
	\, , 
	\label{prdn9}
\end{equation}
which, in the barred quantities $\tilde{g}_{\mu\nu}$ and $\tilde{\phi}$, restores a standard minimally coupled scalar-tensor theory~\cite{Capozziello:2011et}. It is a well-known result~\cite{Nojiri:2010wj} that, on a classical level, the Jordan and Einstein frameworks are dynamically equivalent, while the situation is not definitely clear on a quantum level, see for instance the question concerning the inflationary spectrum in a modified theory of gravity~\cite{Qiu:2014apa}. In this respect, an important physical property of the Jordan frame is the possibility to attribute a fully separate nature to the tensor modes $g_{\mu\nu}$ and $\phi$, which are instead mixed by the conformal transformation in Eq.~(\ref{prdn9}). In fact, as demonstrated in~\cite{Capozziello:2008rq,Moretti:2019yhs}, in a perturbative approach to a Minkowski spacetime, the tensor and scalar modes induce independent oscillation configurations on free particle arrays. This fact leads us to claim that, in a Jordan frame, we can individualize the scalar and tensor components, by which the $f(R)$-gravity is decomposed, as individual quantum modes, say, in the perturbative limit, i.e. they appear as distinct particles. 

In what follows, we will study, in cosmology, the possibility that a massive scalar fluctuation of the scalar field can decay into dark matter constituents. Despite our formulation living on a classical level and the decaying process being summarized by a phenomenological rate, there is no doubt that the physical interpretation of the proposed scenario can be better 
understood in the, here addressed, Jordan frame: we have to consider the decaying scalar massive constituents like real particles, independent from the gravitational field of the expanding Universe. For studies concerning the canonical quantization of modified gravity cosmology in the Jordan frame, see~\cite{DeAngelis:2021afq,DeAngelis:2022qhm,Limongi:2025jsh}.

\section{Dynamics setup}\label{sec2}

We consider a flat and isotropic Universe whose line element reads
\begin{equation}
	ds^2 = -dt^2 + a^2(t) \left( 
	dx^2 + dy^2 + dz^2 \right),
	\label{ht1}
\end{equation}
where $t$ denotes the cosmic (synchronous) time, and $(x, y, z)$ are the Cartesian spatial coordinates. The cosmic scale factor $a(t)$ describes the expansion of the Universe and rescales all spatial lengths. As discussed above, in comparison with GR, metric $f(R)$ gravity, in the \emph{Jordan frame}~\cite{Sotiriou:2008rp,Nojiri:2010wj,Schiavone:2022wvq,Schiavone:2024heb} (see also~\cite{Nojiri:2017ncd}), is characterized by the emergence of a scalar field $\phi$ that is non-minimally coupled to the metric. The potential term $V(\phi)$ associated with this scalar field is directly determined by the original $f(R)$ Lagrangian density, see Eq.~(\ref{prdn5}).

At late times, the $00$-component of the field equations yields the following generalized Friedmann equation:
\begin{equation}
	H^2 \equiv \left( \frac{\dot{a}}{a}\right)^2 
	= \frac{1}{\phi} \left( 
	\frac{\chi}{3}\rho_m - H \dot{\phi} + \frac{V(\phi)}{6} \right),
	\label{ht3}
\end{equation}
where an overdot denotes differentiation with respect to cosmic time, and $\rho_m$ represents the (dark and baryonic) matter energy density, which satisfies the standard conservation equation
\begin{equation}
	\dot{\rho}_m + 3H\rho_m = 0.
	\label{ht4}
\end{equation}
The link between the potential $V(\phi)$ and the evolution of the Hubble parameter $H$ follows from varying the gravitational Lagrangian with respect to $\phi$, giving
\begin{equation}
	\frac{dV}{d\phi} = 12H^2 + 6\dot{H}.
	\label{ht5}
\end{equation}
Combining Eqs.~(\ref{ht3}) and~(\ref{ht5}), one obtains the generalized acceleration equation,
\begin{equation}
	\frac{\ddot{a}}{a} = \dot{H} + H^2 
	= -H^2 + \frac{1}{6}\frac{dV}{d\phi},
	\label{ht6}
\end{equation}
while differentiating Eq.~(\ref{ht3}) and making use of Eq.~(\ref{ht5}) leads to the generalized Klein--Gordon equation for the scalar field:
\begin{equation}
	\ddot{\phi} + 3H\dot{\phi} + \frac{1}{3}\left( \phi \frac{dV}{d\phi} - 2V(\phi) \right) = \frac{\chi}{3}\rho_m.
	\label{ht7}
\end{equation}
Once the Friedmann equation [Eq.~(\ref{ht3})] is specified, Eqs.~(\ref{ht5}) and~(\ref{ht7}) are dynamically equivalent.

\section{Proposed paradigm}\label{sec3}

Following the analyses in~\cite{Montani:2023xpd}, we search for a framework in which the Hubble parameter, the scalar field, and its potential term are all dynamical quantities, while the potential $V(\phi)$—and thus the form of the Lagrangian density $f(R)$—is determined \emph{a posteriori}. To this end, and bearing in mind that the modified gravity must reproduce the present-day vacuum energy density $\rho_{\Lambda}$, we impose the following two conditions:
\begin{equation}
	V(\phi) = 2\chi \rho_{\Lambda} + U(\phi), \qquad 
	6H\dot{\phi} = U(\phi),
	\label{ht8}
\end{equation}
where $U(\phi)$ denotes a generic function to be determined. 

Using the above conditions and switching to the time variable $x = \ln(1+z)$ (where the redshift is defined as $z = 1/a - 1$, with the present-day scale factor normalized to unity), Eqs.~(\ref{ht3}), (\ref{ht5}), and the second relation in Eq.~(\ref{ht8}) can be rewritten as follows. We recall that $\dot{(...)} = -H(...)^{\prime}$, where a prime denotes differentiation with respect to $x$:
\begin{align}
H^2 &= \frac{\chi}{3\phi} \left( 
	\rho_m + \rho_{\Lambda} \right),
	\label{ht9}\\[4pt]
\frac{dU}{d\phi} &= 12H^2 - 3(H^2)^{\prime},
	\label{ht10}\\[4pt]
6H^2\phi^{\prime} &= -U(\phi(x)).
	\label{ht11}
\end{align}
Since $U^{\prime} = \phi^{\prime}\, dU/d\phi$, Eqs.~(\ref{ht10}) and~(\ref{ht11}) yield the following relations:
\begin{equation}
	U(x) = -6\Gamma H e^{-2x}, \qquad 
	\phi^{\prime} = \frac{\Gamma}{H e^{2x}},
	\label{ht12}
\end{equation}
where $\Gamma$ is an integration constant.

Let us now decompose the scalar field as $\phi = \bar{\phi} + \delta\phi$, 
where $\bar{\phi}$ denotes the dominant background component evolving according to the cosmic expansion, and $\delta\phi$ represents a small, time-dependent fluctuation evolving on a much shorter timescale. 
The splitting we adopt is analogous to that discussed in~\cite{Olmo:2005hc}, but we restrict our analysis to the Hubble flow, i.e., we treat both the background and the fluctuating components as homogeneous contributions. 

We further assume that the rapidly varying component $\delta\phi(x)$ decays into dark matter particles, thereby generating a fluctuating matter contribution $\delta\rho_{dm}$ that adds to the standard matter density $\rho_m$. 
The effective dynamics of $\delta\phi$ is obtained by linearizing Eq.~(\ref{ht7}) and introducing a phenomenological decay time $\tau_d \equiv 1/(3\bar{H})$, where $\bar{H}$ is a rate significantly larger than the average value of $H$ during the decay process. 
This leads to the following dynamical equation:
\begin{equation}
	H (H \delta\phi^{\prime})^{\prime} - 3(H + \bar{H}) H \delta\phi^{\prime} 
	+ \mu_{\phi}^2 \delta\phi = \frac{\chi}{3}\, \delta\rho_{dm},
	\label{ht14}
\end{equation}
where the effective mass term (we work in units with $\hbar = 1$) is defined as
\begin{equation}
	\mu_{\phi}^2 \equiv \frac{1}{3} \left( 
	\phi \frac{d^2U}{d\phi^2} - \frac{dU}{d\phi} 
	\right)_{\phi = \bar{\phi}}.
	\label{ht15}
\end{equation}
The quantity $\mu_{\phi}^2$ should remain positive (or vanish) throughout the relevant $x$-interval over which the dynamics is investigated. We emphasize how, a simple calculation shows that, adding to Eq.~(\ref{ht14}) a phenomenological decaying rate implies that in Eq.~(\ref{ht10}) we correspondingly add, on the right-hand-side, the term $9\bar{H}\delta \phi^{\prime}/\bar{\phi}$.

The fluctuation in the dark matter component must satisfy the following equation, which ensures energy balance during the decay process:
\begin{equation}
	\delta\rho_{dm}^{\prime} - 
	3\,\delta\rho_{dm} = 
	\frac{1}{\chi}\left[ 
	-3\bar{H}H\bar{\phi}^{\prime} 
	+ H(H\bar{\phi}^{\prime})^{\prime}
	\right] \delta\phi^{\prime},
	\label{ht16}
\end{equation}
where the prime denotes differentiation with respect to $x$.

It is important to note that, when introducing the decay rate $\bar{H}$ of the perturbed scalar field, the field equations governing $H^2$ and $\dot{H}$ must remain unchanged at first order in $\delta\phi$. Within this approximation scheme, Eqs.~(\ref{ht12}) take the following form:
\begin{equation}
	\bar{U}(x) + \bar{U}^{\prime}\frac{\delta\phi}{\bar{\phi}^{\prime}} 
	= -6\Gamma H e^{-2x} 
	\exp\!\left[ -\frac{9}{2}\left(\frac{\bar{H}}{H} \right)^2
	\int_0^x \frac{dy\,\delta\phi}{H\bar{\phi}} \right],
	\label{ht17}
\end{equation}
where $\bar{U}(x) \equiv U(\bar{\phi}(x))$, and
\begin{equation}
	\bar{\phi}^{\prime} + \delta\phi^{\prime} 
	= \frac{\Gamma}{H e^{2x}}
	\exp\!\left[ -\frac{9}{2}\left(\frac{\bar{H}}{H} \right)^2
	\int_0^x \frac{dy\,\delta\phi}{H\bar{\phi}} \right].
	\label{ht18}
\end{equation}
Finally, the generalized Friedmann equation~(\ref{ht9}) can now be restated as
\begin{equation}
	H^2 = \frac{\chi}{3\bar{\phi}} 
	\left[
	\left( \rho_m + \rho_{\Lambda} \right)
	\left( 1 - \frac{\delta\phi}{\bar{\phi}} \right) 
	+ \delta\rho_m
	\right],
	\label{ht19}
\end{equation}
where only linear contributions in the perturbed quantities have been retained.

\section{Reduced dimensionless model}\label{sec4}

We can now significantly simplify the present model by considering the regime in which the decay rate $\bar{H}$ is much larger than both the average value of the Hubble parameter $H$ and the scalar field mass $\mu_{\phi}$ in the late Universe. 
Also taking into account the smallness of the Einstein constant, Eq.~(\ref{ht14}) reduces to
\begin{equation}
	(H\delta\phi^{\prime})^{\prime} 
	- 3\bar{H}\delta\phi^{\prime} = 0.
	\label{ht20}
\end{equation}
This equation admits the solution
\begin{equation} 
	\delta\phi = \delta\phi_0 
	\exp\!\left[
	3\xi_0 \int_0^x \frac{dy}{E(y)}
	\right],
	\label{ht21}
\end{equation}
where $\delta\phi_0 \equiv \delta\phi(0) > 0$ is an integration constant (a second, nonphysical one has been set to zero), and we define $\xi_0 \equiv \bar{H}/H_0$ with $H_0 \equiv H(x=0)$. 
We also introduce the normalized expansion rate $E(x) \equiv H(x)/H_0$, for which $E(0)=1$. 
In terms of these quantities, Eq.~(\ref{ht19}) can be recast as
\begin{equation}
	E^2 = \frac{1}{\bar{\phi} + \delta\phi}
	\left[
	\left( \Omega_m^0 e^{3x} + 1 - \Omega_m^0 \right) 
	+ \delta\Omega_{dm}
	\right],
	\label{ht23}
\end{equation}
where we have used the solution of Eq.~(\ref{ht4}), $\rho_m = \rho_m^0 e^{3x}$, together with the standard definitions 
$\Omega_m^0 \equiv \chi\rho_m^0 / (3H_0^2)$ and 
$\delta\Omega_{dm} \equiv \chi\delta\rho_{dm} / (3H_0^2)$. 
By setting $\bar{\phi}(0) = 1$ to recover GR today (with $|\delta\phi_0| \leq 10^{-7}$~\cite{Hu:2007nk}), we immediately obtain
\begin{equation}
	\delta\Omega_{dm}(0) = \delta\phi_0.
	\label{ht25}
\end{equation}

It is straightforward to verify that Eq.~(\ref{ht16}) admits the following normalized leading-order solution:
\begin{equation}
	\delta\Omega_{dm} = C 
	- \frac{3\xi_0}{\sqrt{\bar{\phi}}} 
	\sqrt{\Omega_m^0 e^{3x} + 1 - \Omega_m^0}\,
	\bar{\phi}^{\prime}\delta\phi,
	\label{ht26}
\end{equation}
where $C$ is an integration constant. 
From Eq.~(\ref{ht18}), we also obtain
\begin{equation}
	\bar{\phi}^{\prime} = 
	\frac{\gamma_0}{
	e^{2x}\sqrt{\bar{\phi}}
	\sqrt{\Omega_m^0 e^{3x} + 1 - \Omega_m^0}}
	+ \mathcal{O}(\delta\phi),
	\label{ht27}
\end{equation}
where $\gamma_0 \equiv \Gamma / H_0$.

We now have all the necessary ingredients to construct the cosmological dynamics associated with the proposed dark energy--dark matter interaction scenario. 
In what follows, we focus on the case in which the considered $f(R)$ model introduces only a small deviation from GR.

\section{Model dynamics near GR}\label{sec5}

The near-GR limit results characterized by the following conditions:
\begin{equation}
	\bar{\phi} \equiv 1, \qquad
	\bar{U}(1) \simeq 0, \qquad 
	\left.\frac{d\bar{U}}{d\bar{\phi}}\right|_{\bar{\phi}=1} \simeq 0, \qquad
    \mu_{\phi}^2 = \frac{1}{3}\left.\frac{d^2\bar{U}}{d\bar{\phi}^2}\right|_{\bar{\phi}=1} \ge 0.
	\label{xx}
\end{equation}
Within the framework outlined above, these conditions correspond to setting 
$\Gamma \simeq 0$, i.e., $\gamma_0 \simeq 0$. Under this approximation, the normalized Hubble expansion rate takes the form
\begin{equation}
	E^2(x) = 
	\frac{1}{1 + \delta\phi_0\, F(x)}
	\left( 
	\Omega_m^0 e^{3x} 
	+ 1 - \Omega_m^0 
	+ \delta\phi_0
	\right),
	\label{htt7}
\end{equation}
where $F(x)$ satisfies
\begin{equation}
	F^{\prime} = \frac{3\xi_0 F(x)}{E(x)}, \qquad F(0) = 1.
	\label{ht28}
\end{equation}

In this approximation, the dark matter fluctuation $\delta\Omega_{dm}$ is, \emph{de facto}, created instantaneously near $x = 0$. 
However, it is straightforward to see that, due to the normalization condition, this term can exist only if $\delta\phi \neq 0$, and in particular $\delta\phi_0 \neq 0$.

We now intend to investigate if the proposed dynamical scenario is, actually, equivalent when restated into an Einstein frame. In fact, while we consider as the most appropriate formulation to our massive particle decaying process the one emerging in the 
Jordan frame (where the scalar mode 
has a well-defined physical entity, as discussed in Sec.\ref{sec00}), the 
formally classical nature of our 
formulation suggests that a possible equivalence exists. Since the decaying process is phenomenologically described via an effective rate $\bar{H}$, 
the question about the equivalence should be properly formulated in asking 
which is the phenomenological modification in the Einstein frame able to guarantee the equivalence of the two approaches. 

Being well-established that the two frames are certainly equivalent on 
a classical (non-modified) paradigm \cite{Sotiriou:2008rp,2011PhR...509..167C,deHaro:2023lbq}, we can reliably conclude that, for what concerns the background dynamics of our model, no equivalence question arises. Then the investigation must be restricted to linearized perturbation dynamics only. The basic equation, governing the evolution of the fluctuation $\delta\phi$ is easily identified in 
Eq. (\ref{ht14}), which is, however, discussed in the limit in which the decaying rate dominates the contribution of the 
Universe expansion, as well as the 
massive term. In other words, the 
real dynamical equation we addressed in our study is Eq. (\ref{ht20}). To understand how such an equation is 
formulated in the Einstein frame, we can make use of the conformal transformation in Eq. (\ref{prdn9}). In the linearized 
dynamics, it is immediate to 
recognize the transformation
\begin{equation}
	\phi = 1 + \delta \phi \simeq 
	\exp \Big[\sqrt{\frac{2\chi}{3}}\tilde{\phi}\Big] \simeq 
	1 + \sqrt{\frac{2\chi}{3}}\tilde{\delta \phi}
	\, , 
	\label{eq2}
\end{equation}
so that, we obtain a direct proportionality between $\delta \phi$ and 
$\tilde{\delta \phi}$. We underline that we are focusing our attention to the near-GR limit for which 
the background scalar field is taken 
equal to unity in the Jordan frame. 
Furthermore, we easily get that 
\begin{equation}
	H=\tilde{H} - \frac{1}{2}
	\sqrt{\frac{2\chi}{3}} \dot{\delta \phi} 
	\, 
	\label{eq3}
\end{equation}
and, passing to the time variable $x$, we finally get
\begin{equation}
	H = \frac{\tilde{H}}{1 + 
	\frac{1}{2}\sqrt{\frac{2\chi}{3}}\tilde{\delta \phi}^{\prime}}
	\, .
	\label{eq4}
\end{equation}

Substituting Eqs. (\ref{eq2}) and (\ref{eq4}) into Eq. (\ref{ht20}) and linearizing in $\tilde{\delta \phi}$, we see how this equation retains the same form in the Einstein frame: the dynamics is preserved equivalent between the Jordan and Einstein frames, 
and also the minimally coupled scalar field, emerging in the latter, is subjected to a phenomenological decaying process with the same phenomenological rate. We conclude by stressing how the 
non-linear corrections in $\delta \phi$ and $\tilde{\delta \phi}$ do not 
concern our analysis, simply because the scenario of decaying massive free particles would be violated.

\section{Methodology of data analysis}\label{sec5bis}

The resulting model in the near-GR limit thus corresponds to a modified version of $\Lambda$CDM, characterized by four free parameters: 
$H_0$, $\Omega_m^0$, $\delta\phi_0$, and $\xi_0$. 
We further assume a very rapid decay of the scalar mode, $\bar{H} \gg H_0$, which implies $\xi_0 \gg 1$, while respecting the bound $|\delta\phi_0| \leq 10^{-7}$. 
From the resulting expression for the Hubble parameter, the most consistent phenomenological interpretation is that proposed in~\cite{Montani:2024pou}: the function $\delta\phi$ is taken to vanish identically once the Hubble parameter approaches the DESI+Planck central value curve. 
This prescription introduces a discontinuity in the derivative of $\delta\phi$, whose sharpness quantifies the dark matter creation process as a transitional phase in the late-time cosmic evolution. 
The subsequent data analysis is performed within this approximation scheme, where $\delta\phi$ is effectively switched off after the transition.

To assess the viability of our modified $\Lambda$CDM framework, hereafter referred to as the MG (Modified Gravity) model, we confront its predictions with cosmological data using \texttt{SimpleMC}~\cite{simplemc}, a sampler designed for parameter inference from cosmological background observations. 
Since the dark energy decay process is expected to occur at late times, $0.001 < z < 1.0$ (i.e., $0.0004 < x < 0.3$), we restrict our analysis to background-relevant parameters. 
From the standard model, we retain the matter density parameter $\Omega_m^0$, the dimensionless Hubble constant $h$ (defined as $h = H_0 / 100$), and the baryon density $\Omega_b h^2$. 
Our model introduces two additional parameters, $\delta\phi_0$ and $\xi_0$, which determine, respectively, the onset and rate of the matter creation process. 
For comparison, we also perform a parallel analysis of the baseline $\Lambda$CDM model, characterized by the three standard parameters above.

For the priors used, we select agnostic ones (flat) for every parameter. The specific range is: $h=[0.4,0.9]$, $\Omega_m^0 = [0.1,0.5]$, $\Omega_bh^2 = [0.020, 0.025]$, $\delta\phi_0=[10^{-14},10^{-4}]$, and $\xi_0=[1,200]$ (for the sake of completeness, those last two priors are extended also into the physically forbidden region). The sampling method chosen for this work is the one referred to as dynamic nested sampling~\cite{Feroz:2007kg,Feroz:2013hea}, which in \texttt{SimpleMC} is implemented by using the \texttt{dynesty} library~\cite{Speagle:2019ivv}. To decide the number of live points, we follow the general rule~\cite{dynesty} of using $50\times ndim$, where the $ndim$ corresponds to the number of free parameters (also referred to as \textit{dimensionality}).

The datasets used to make the parameter inference procedure are as follows: 
\begin{itemize}
    \item CC: a compilation of 15 Cosmic Chronometer (CC) measurements, provided together with their full covariance matrix~\cite{Moresco:2020fbm};
    \item SN: the PantheonPlus sample of Type Ia Supernovae (SN)~\cite{Scolnic:2021amr,Brout:2022vxf}, consisting of 1701 light curves from 1550 unique SN, used to determine the distance modulus;
    \item BAO: the most recent (DR2) Baryon Acoustic Oscillation (BAO) measurements from DESI~\cite{DESI:2025zgx,DESI:2025fii,DESI:2025qqy};  
    \item SH0ES: the distance-ladder calibration applied to the PantheonPlus SN sample~\cite{Riess:2021jrx}.
\end{itemize}
Given that they are considered independent from each other, we can go ahead and treat the overall $\chi^2_{total}$ as
\begin{equation}
    \chi^2_{total} = \chi^2_{\tt CC} + \chi^2_{\tt BAO} + \chi^2_{\tt SN} + \chi^2_{\tt SH0ES}
\label{chi2}
\end{equation}
where, for each dataset, we have
\begin{equation}
    \chi^2_{\rm data}= (d_{i,m}-d_{i,{\rm obs}})C_{ij,{\rm data}}^{-1}(d_{j,m}-d_{j,{\rm obs}}),
\end{equation}
and here $d_{m}$ are our model predictions, $d_{\rm obs}$ are the observables, and $C_{\rm data}$ is the associated covariance matrix.

In addition to the $\chi^2$ analysis, we perform a Bayesian model comparison by computing the logarithm of the Bayesian evidence, $\ln \mathcal{Z}$. According to Bayes’ theorem, for each model $\mathcal{M}_i$ with parameters $\Theta$, the posterior distribution is given by
\begin{equation}
P(\Theta|D,\mathcal{M}_i) = \frac{\mathcal{L}(D|\Theta,\mathcal{M}_i)\,\pi(\Theta|\mathcal{M}_i)}{\mathcal{Z}_i},
\label{eq:bayes_theorem}
\end{equation}
where $\mathcal{L}$ denotes the likelihood, $\pi$ represents the prior, and the Bayesian evidence $\mathcal{Z}_i$ is defined as
\begin{equation}
\mathcal{Z}_i = \int \mathcal{L}(D|\Theta,\mathcal{M}_i)\,\pi(\Theta|\mathcal{M}_i)\,{\rm d}\Theta.
\label{eq:bayesian_evidence}
\end{equation}

The relative strength of two competing models, $\mathcal{M}_i$ and $\mathcal{M}_j$, is quantified by the Bayes factor $B_{ij} = \mathcal{Z}_i / \mathcal{Z}_j$. In our analysis, we compute $\ln B_{\Lambda \text{CDM},i} = \ln \mathcal{Z}_{\Lambda \text{CDM}} - \ln \mathcal{Z}_i$, where positive values indicate a preference for $\Lambda$CDM. Following the empirical Jeffreys scale~\cite{Kass:1995loi}, $|\ln B_{\Lambda \text{CDM},i}| \lesssim 1$ is considered inconclusive, $1 \lesssim |\ln B_{\Lambda \text{CDM},i}| \lesssim 3$ indicates moderate evidence, $3 \lesssim |\ln B_{\Lambda \text{CDM},i}| \lesssim 5$ corresponds to strong evidence, and values above $|\ln B_{\Lambda \text{CDM},i}| = 5$ are regarded as decisive.

\section{Results and discussion}\label{sec6}

We now discuss the results obtained from the parameter inference procedure (we emphasize that, in this section, we revert to the redshift variable $z$ to facilitate a more direct comparison with the data and the standard literature). Using the datasets described in the previous section, we obtained the mean values and their corresponding $1\sigma$ uncertainties for the parameters, which are reported in Table~\ref{tabla_evidencias}.
\begin{table*}[th!]
\footnotesize
\scalebox{1.2}{%
\begin{tabular}{cccccccc} 
\cline{1-8}\noalign{\smallskip}
\vspace{0.15cm}
Model & $H_0$ & $\Omega_bh^2$ & $\Omega^{0}_{m}$ &  $\log(\delta\phi_0)$ &  $\xi_0$  & $\ln B_{\Lambda \text{CDM},i}$  &  $-2\Delta\ln \mathcal{L_{\rm max}}$ \\
\hline
\vspace{0.15cm}
$\Lambda$CDM  & $68.78\pm 0.52$ &  $0.0222\pm 0.0004$  &  $0.305\pm 0.007$ & $-$ & $-$ & $-$  &  $-$  \\
\vspace{0.15cm}
MG & $70.27\pm 0.61$  & $0.0218\pm 0.0004$  & $0.330\pm 0.008$ & $-4.2^{+0.18}_{-0.32}$ &  $< 5.12$  & $-0.16\pm0.23$  & $-9.73$ \\
\hline
\hline
\end{tabular}}
\caption{Summary of the mean values and standard deviations of the parameters for the $\Lambda$CDM and MG models using the data combination CC+SN+BAO+SH0ES. The last two columns correspond to the natural logarithm of the Bayes factor, $\ln B_{\Lambda \text{CDM},i}$, which, if positive, indicates a preference for $\Lambda$CDM, and to $-2\Delta\ln \mathcal{L}_{\rm max} \equiv -2\ln(\mathcal{L}_{\rm max,\Lambda \text{CDM}} / \mathcal{L}_{\rm max},_i)$, which, if positive, indicates that $\Lambda$CDM provides a better fit to the data.}
\label{tabla_evidencias}
\end{table*}

\begin{figure}[ht!]
    \centering
    \includegraphics[scale=0.65]{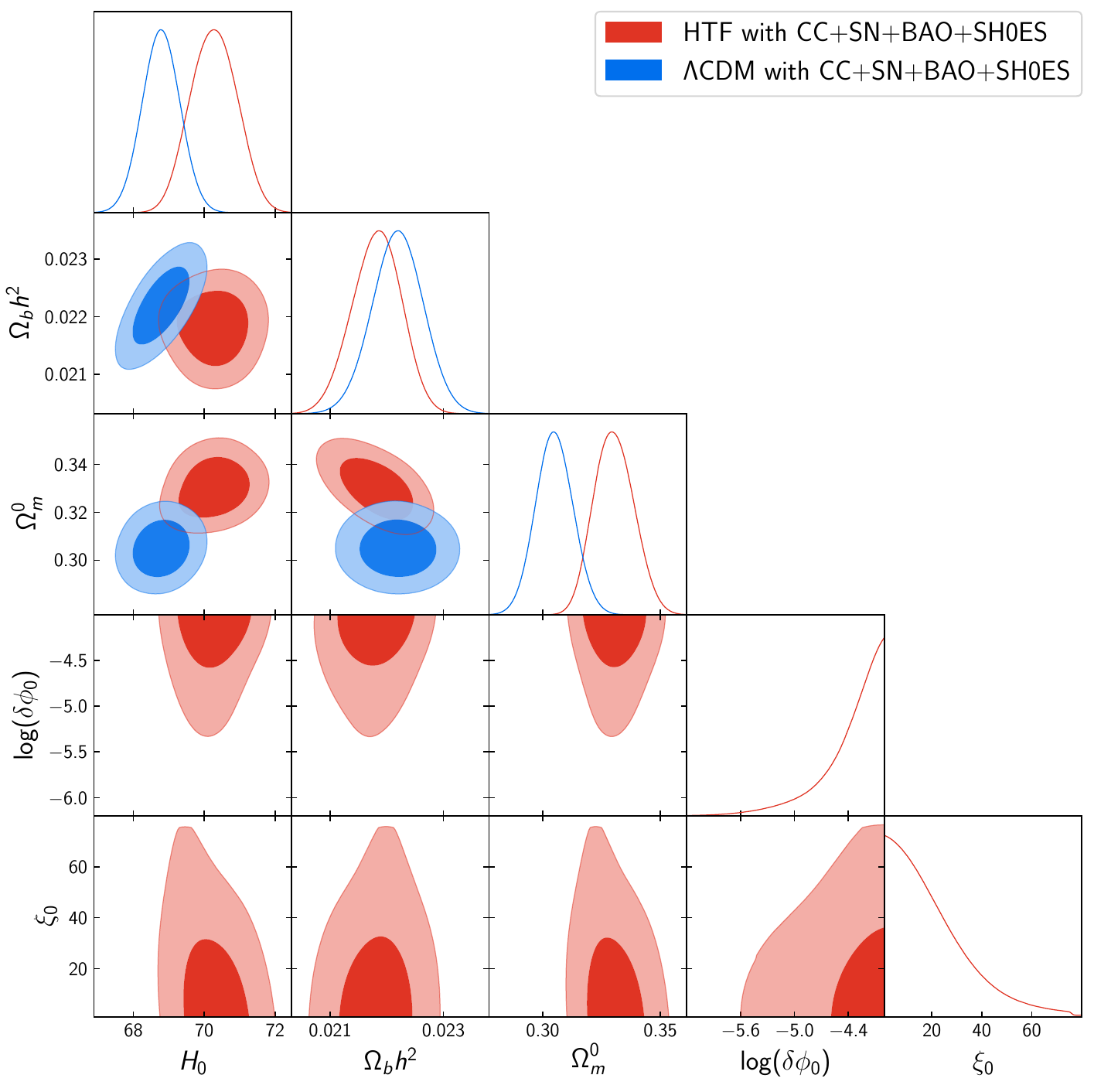} 
    \caption{Triangle plot showing the 1D and 2D marginalized posteriors from the parameter inference analysis. Since the parameter $\delta\phi_0$ spans several orders of magnitude, we show its logarithm instead.}
    \label{fig:result_triangleplot}
\end{figure}

The marginalized 1D and 2D posteriors are shown in Fig.~\ref{fig:result_triangleplot}, where we have replaced $\delta\phi_0$ with its logarithm, $\log(\delta\phi_0)$. For the MG model, the inferred value of $H_0 = 70.27 \pm 0.61$~km\,s$^{-1}$\,Mpc$^{-1}$ is higher than that obtained for $\Lambda$CDM. While the SH0ES prior is included in both analyses, the MG model provides a better overall fit to the data thanks to its greater flexibility. The additional parameters introduced in the MG model do not show significant correlations with $H_0$ or $\Omega_m$, indicating that the observed shift in these quantities stems from the model’s extended parameter space rather than from parameter degeneracies. The model also yields an increase of about $\sim10\%$ in the present matter density of the Universe. It is easy to verify that $q_0 = 0.49 \pm 0.3$ for our model, which lies slightly below the $2\sigma$ level from the standard model's expected value of $0.55$.

There are two key quantities required for model comparison: the minimum $\chi^2$, which indicates how well a model fits the data, and the Bayesian evidence, $\ln Z$, which quantifies whether one model is preferred over another. The values of $\chi_{\tt min}^2$ and $\ln Z$ are $1425.64$ ($\chi^2_{\tt CC}=16.24$, $\chi^2_{\tt SN}=1402.18$, $\chi^2_{\tt BAO}=5.43$, $\chi^2_{\tt SH0ES}=1.79$) and $-726.32 \pm 0.18$ for our MG model, and $1435.37$ ($\chi^2_{\tt CC}=14.56$, $\chi^2_{\tt SN}=1405.34$, $\chi^2_{\tt BAO}=11.11$, $\chi^2_{\tt SH0ES}=4.36$) and $-726.16 \pm 0.17$ for $\Lambda$CDM, respectively. Since the datasets are independent, we have separated the total $\chi_{\tt min}^2$ into its individual contributions as shown in Eq.~(\ref{chi2}). Our MG model shows an improvement of approximately $10$ in $\chi^2$ compared to $\Lambda$CDM. However, as it includes two additional free parameters beyond the three of $\Lambda$CDM, the Bayesian evidence remains inconclusive, with a difference of less than unity between the two models, according to the empirical Jeffreys scale~\cite{Kass:1995loi}.

\begin{figure}[ht!]
    \centering
    \includegraphics[scale=0.7]{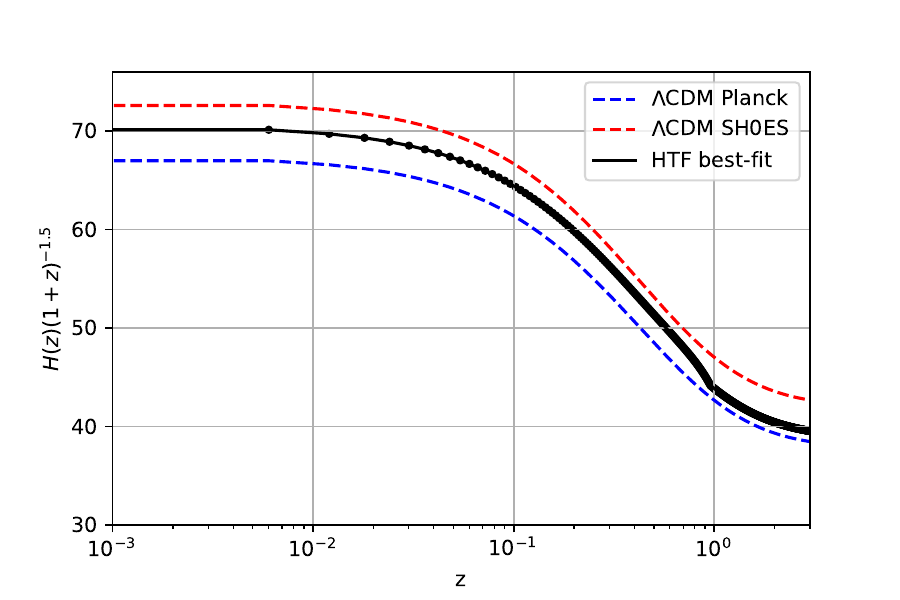} 
    \includegraphics[scale=0.7]{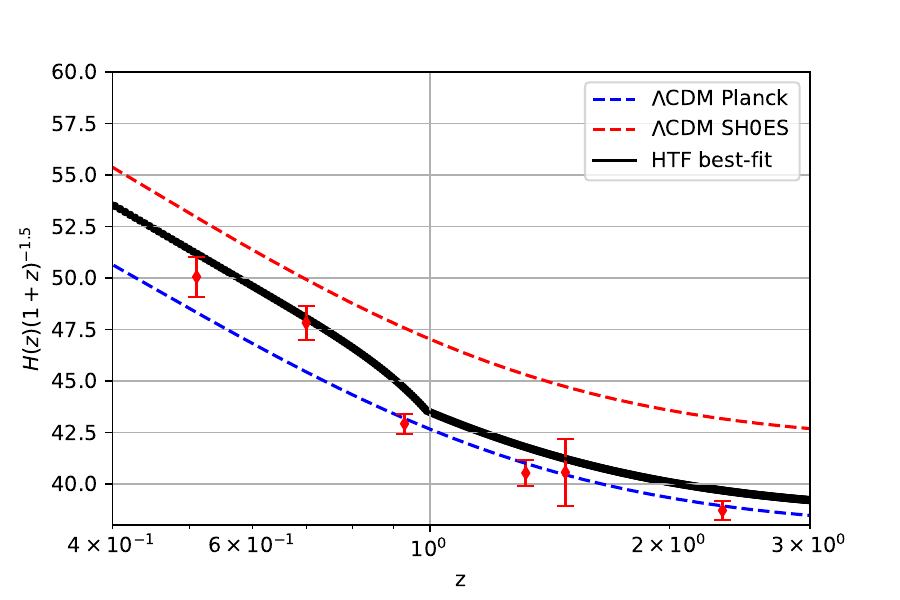} 
    \caption{Plots showing the normalized $H(z)$. \textit{Upper:} Normalized $H(z)$ in the redshift range $0<z<3$. \textit{Lower:} Normalized $H(z)$ in the range $0.4<z<3.0$, including the line-of-sight BAO data from DESI DR2.}
    \label{fig:result_bestfit}
\end{figure}

Examining the individual contributions to the total $\chi^2_{\rm min}$, it becomes evident that the improvement in the fit arises mainly from the SN data ($\Delta\chi^2_{\tt SN}=3.16$), the BAO data ($\Delta\chi^2_{\tt BAO}=5.68$), and the SH0ES calibration ($\Delta\chi^2_{\tt SH0ES}=2.57$). The SH0ES contribution is not unexpected, as the phenomenological nature of our model produces a ``sudden rise'' in $H(z)$ at late times, favoring values of $H_0$ higher than those inferred from Planck’s CMB data. More interesting, however, are the SN and BAO contributions, which merit closer inspection. 

In Fig.~\ref{fig:result_bestfit}, we show the evolution of $H(z)/(1+z)^{1.5}$ for the MG model using the best-fit parameters. The model predicts that the decay into dark matter occurs around $z \sim 1$, offering a possible explanation for the BAO preference. The DESI BAO dataset covers the redshift range $0.3 < z < 2.3$, with six data points at $z > 1$ (from the ELG2, QSO, and Ly$\alpha$ tracers). When the line-of-sight BAO measurements are included, it becomes evident that the BAO data below $z \lesssim 1$ are mainly responsible for the improved fit, as shown in the lower panel of Fig.~\ref{fig:result_bestfit}.  

This interpretation is further supported by the DESI analysis presented in~\cite{DESI:2025zgx}, particularly in the discussion around their Fig.~9. There it is shown that line-of-sight BAO measurements at $z>1.1$ exhibit degeneracies distinct from those at $z<1.1$. Moreover, when analyzed separately (in combination with BBN constraints), the BAO data below $z<1.1$ display a clear preference for higher values of both $\Omega_m^0$ and $H_0$ (for a related effect in binned quasar data, see~\cite{Dainotti:2024aha}). Since our MG model phenomenologically induces a transition near $z\sim1$, shifting from lower to higher values of $\Omega_m^0$ and $H_0$, it naturally aligns with this behavior.  

Therefore, the improved $\chi^2$ for BAO is not accidental but instead reflects that the MG dynamics reproduce a feature already present in the data. This preference is also consistent with the trend indicated by the SH0ES calibration, providing a coherent explanation for why the model performs better than $\Lambda$CDM and yields a slightly higher $H_0$. The concordance between these independent datasets underscores the physical relevance of the model’s late-time transition and reinforces the viability of MG-like scenarios as alternatives to the $\Lambda$CDM model.

\section{Conclusions}

We implemented a cosmological model based on metric $f(R)$ gravity in the Jordan frame, characterized by the peculiar feature that the non-minimally coupled scalar field~\cite{Olmo:2005zr} is decomposed into a background component (set to unity to recover GR) and a rapidly varying fluctuation. This separation relies on the massive nature of the fluctuation, which undergoes a rapid decay into the dark matter component of the Universe.

This scenario represents, \emph{de facto}, a cosmological framework in which a fraction of the dark energy component is transferred, through a phase transition, to dark matter particles. The decay process is modeled phenomenologically through an effective decay rate and naturally motivates the construction of a fundamental interaction cross section between the massive scalar fluctuations and the dark matter constituents. A formulation of this kind would, however, depend strongly on the specific quantum field theory assumed at the fundamental level. Here, instead, our aim is to explore the potential of this scenario to mitigate the Hubble tension, focusing on its most general and model-independent features.

Phenomenologically, this scenario can be interpreted as a dark matter particle creation model in the near-GR limit. It, therefore, constitutes a minimal modification of the standard $\Lambda$CDM framework, sharing its background parameters while introducing two additional ones: an integration constant, $\delta\phi_0$, and a normalized interaction rate, $\xi_0$. The model retains most of the background dynamics of $\Lambda$CDM but, depending on the values of these new parameters, it can alleviate one of the most persistent challenges to the standard model: the Hubble tension.

By constraining the parameters of the model with the latest cosmological background data and comparing the results with those obtained for the vanilla $\Lambda$CDM model, we found an overall improvement in the fit. Although the Bayesian evidence remains inconclusive and does not favor the MG model over $\Lambda$CDM, this outcome is encouraging given that the former includes two additional free parameters. A detailed analysis of the $\chi^2$ contributions indicates that the improvement is mainly driven by the SH0ES calibration and the DESI BAO measurements, particularly in the redshift range $z < 1$.  

While the recovered parameter values do not fully resolve the Hubble tension, the results are promising. They highlight the ability of a decaying dark energy cosmology to achieve a better reconciliation with low-redshift BAO data than $\Lambda$CDM, while remaining consistent with other background probes. This provides a strong motivation for further investigation of interacting dark sector scenarios, particularly their implications for structure formation and perturbative dynamics, which may yield additional signatures distinguishing them from the standard model. This work represents a preliminary background-level analysis demonstrating the potential of the model. A comprehensive study including CMB data will be necessary to fully assess its viability, which we leave for future work.

\section*{ACKNOWLEDGMENTS}
EDV is supported by a Royal Society Dorothy Hodgkin Research Fellowship.
We acknowledge the IT Services at The University of Sheffield for the 
provision of services for High Performance Computing. LAE acknowledges partial financial support from the T\"{u}rkiye Bilimsel ve Teknolojik Ara\c{s}t{\i}rma Kurumu (T\"{U}B\.{I}TAK, Scientific and Technological Research Council of T\"{u}rkiye) through grant no.\ 124N627.
This article is based upon work from the COST Action CA21136 - ``Addressing observational tensions in cosmology with systematics and fundamental physics (CosmoVerse)'', supported by COST - ``European Cooperation in Science and Technology''.


\end{document}